\documentclass[12pt]{iopart}
\usepackage[dvips]{epsfig}

\newcommand{\figwidth}{0.98\columnwidth}
\newcommand{\vect}[1]{\mathbf{#1}}
\begin{document}
\title{Formation of the $S=1$ paramagnetic centers in the bond-diluted spin-gap magnet.}

\author{V N Glazkov$^{1,2}$, Yu V Krasnikova$^{1,2}$, D H\"{u}vonen$^{3,4}$, A Zheludev$^{4}$}

\address{$^1$ P.L.Kapitza Institute for Physical Problems RAS,
Kosygin str. 2, 119334 Moscow, Russia}

\address{$^2$ Moscow Institute of Physics and Technology, 141700
Dolgoprudny, Russia}

\address{$^3$ National Institute of Chemical
Physics and Biophysics, Akadeemia tee 23, 12618 Tallinn, Estonia}

\address{$^4$ Neutron Scattering and Magnetism, Laboratory for
Solid State Physics, ETH Z\"{u}rich, 8006 Z\"{u}rich, Switzerland}

\ead{glazkov@kapitza.ras.ru}

\begin{abstract}
Electron spin resonance experiment reveals that non-magnetic bond doping of the spin-gap magnet (C$_4$H$_{12}$N$_2$)Cu$_2$Cl$_6$ (abbreviated PHCC) results in the formation of $S=1$ paramagnetic centers that dominate low-temperature ESR response. We have followed evolution of this signal with doping impurity content and have found that these centers concentration is quadratic over the impurity content. We also observe coexistence of the ESR responses from these local centers and from delocalized triplet excitations over certain temperature range.

\end{abstract}

\date{\today}



\submitto{JPCM}

\section{Introduction}
Spin-gap magnets have been actively studied during the last decades.
These systems are characterized by the presence of a nonmagnetic
singlet ($S=0$) ground state separated from the excited triplet
($S=1$) states by an energy gap $\Delta$ of exchange origin. One
of the most striking features of these systems is that despite
strong exchange coupling they do not order magnetically down to
lowest temperatures.

Stability of the singlet ground state against various
perturbations: additional spin-spin interactions, external
magnetic field, pressure and introduction of impurities is a long discussed problem.
Magnetic field lifts degeneracy of the triplet states and splits
triplet sublevels. At the critical field $H_c=\Delta/(g\mu_B)$
energy of one of the triplet sublevels becomes lower than the
energy of the singlet ground state. At this point phase transition
occurs which was actively discussed in the context of
Bose-Einstein condensation of
magnons \cite{giamarchi,zapf-revmodphys}.

Impurities can substitute magnetic ions depleting magnetic system
(site-doping) or alternate superexchange pathes (bond-doping).
Impurity ions vary material properties locally disturbing and
sometimes even destroying the spin-gap state of the parent
material. This leads to the variety of known effects. Formation of
the impurity induced magnetic order was observed in a spin-Peierls
magnet CuGeO$_3$ \cite{cugeo3-Zn,cugeo3-imps}, Haldane magnet
 \cite{pbnio-AFM} or a prototypical dimer magnet
TlCuCl$_3$ \cite{TlCuCl3-doped1,TlCuCl3-doped2}. Randomization of
the exchange couplings through the bond doping allows for
possibility of realization of a Bose-glass
state \cite{glassy,bg-exp} or a universal low-energy physics of
a random spin chain \cite{random1,random2,random-exp}.

Violation of the gapped state in the impurity vicinity can lead to a formation of a multi-spin defects with total non-zero spin. These defects (or spin clusters) can demonstrate unusual properties: Spin clusters in Ni-doped spin-Peierls magnet CuGeO$_3$ demonstrated unusual (as small as 1.4) and highly-anisotropic $g$-factor \cite{cugeo3-Ni,cugeo3-Ni2,cugeo3-Ni3}. Spin clusters formed at the ends of the chain fragment of a Haldane $S=1$ magnet are shown both theoretically \cite{spin-half-haldane-theory} and experimentally \cite{spin-half-haldane-exp1,spin-half-haldane-exp2,spin-half-haldane-exp3} to carry total spin $S=1/2$ each.

Our recent study \cite{phcc-doped} demonstrated that bond-doping in
the quasi-2D spin-gap magnet (C$_4$H$_{12}$N$_2$)Cu$_2$Cl$_6$
(abbreviated PHCC) leads to the formation of yet
another type of defect: a local $S=1$ defect. This is a surprising outcome, as magnetic ions of PHCC are antiferromagnetically coupled $S=1/2$ Cu$^{2+}$ ions. Effect
of bond-doping in this case can be interpreted as variation of the effective potential acting on the triplet quasiparticles (which are freely moving quasiparticles in
a pure system). As there is always a bound state in a 2D potential well, one can expect that some of triplet excitations could be trapped in the defect vicinity if this effective potential is attractive. In the case of bromine-substituted PHCC this
potential turns out to be not only attractive, but strong enough to yield a
low-lying bound state. In the present report we demonstrate new
experimental findings that provides additional proof of this
effect.

\section{Samples and experimental details}

Samples of Br-substituted PHCC
(C$_4$H$_{12}$N$_2$)(Cu$_2$Cl$_{6(1-y)}$Br$_{6y}$) were grown
from the solution as described in Ref. \cite{tatiana-growth}.
We used samples from the same set as were used in
Refs. \cite{dan,phcc-pure,phcc-doped,prb-dan-doped,prb-dan-doped-magnons}.
We will use nominal Br concentration $x$ for sample identification
as it was used in earlier papers.

PHCC crystallizes in a triclinic structure, as-grown crystals are
elongated along the $c$-axis and has a well developed plane
orthogonal to $a^*$ direction. As PHCC became almost non-magnetic
at low temperatures large crystals (approx. $2\times3\times6$mm$^3$) were used which limited
mounting possibilities in our experimental setup. To compare data
with earlier work, we mounted samples in the $\vect{H}||a^*$
orientation (field applied orthogonal to the natural plane of the
crystal).

Magnetic ions Cu$^{2+}$ form planes in the $(ac)$ crystallographic
plane of PHCC. Inelastic neutron scattering studies \cite{ins,stone}
indicates presence of 6 to 8 relevant in-plane exchange bonds.
Geometry of the exchange bonds can be envisioned as a set of
strongly coupled ladders with dominating in-rung exchange and some
frustrating couplings. Thus, a dimer Cu$_2$Cl$_6$ with the
strongest antiferromagnetic exchange coupling (bond 1 in terms of
Ref. \cite{stone}) can be considered as a building block of
the PHCC magnetic structure. Intra-dimer exchange is mediated by
two halogen ions.

Up to nominal bromine concentration of 12\% crystals remain
single-phase. Detailed structural analysis of Ref. \cite{dan}
have demonstrated that occupancy of different halogen sites
differs from the nominal concentration. However, actual occupancy
of all sites is linear with nominal concentration of the bromine.
Average actual bromine concentration is \cite{dan} $y=0.63 x$.

ESR absorption was studied using set of homemade transmission type
spectrometers in the temperature range from 400mK to 20K.
Microwave power transmitted through the microwave cavity with the
sample depends upon imaginary part of susceptibility as
$P_{trans}(H)\propto 1/(1+A\chi''(H))^2$, here factor $A$ depends
upon experimental conditions (cavity Q-factor and sample position
inside the cavity) and is the same within given experiment. ESR
absorption can be scaled with static magnetization data since
paramagnetic resonance integral intensity $I=\int \chi''(H) dH$ is
proportional to the static susceptibility.

ESR is a sensitive tool to probe low-energy dynamics of a magnet.
In the case of a $S=1$ magnetic ion or a triplet excitation of the
spin gap magnet anisotropic interactions lead to the zero-field
splitting of the triplet level. This splitting is described by the
term $DS_z^2$ in the spin Hamiltonian in the axial case and by a
terms $DS_z^2+E(S_x^2-S_y^2)$ in a general case. In the case of
weak splitting magnetic field direction is a quantization axis and
effective anisotropy constants $D$ and $E$ are orientation
dependent. We will assume axial anisotropy (i.e., $E=0$) in the
qualitative discussions of this paper.

As a result of zero-field splitting resonance fields in the
fixed-frequency experiment differ for $\left| S_z=+1\right \rangle
\rightarrow \left| S_z=0\right \rangle$ and $\left| S_z=-1\right
\rangle \rightarrow \left| S_z=0\right \rangle$ transitions as
$H_{1,2}=H_0\pm D/(g\mu_B)$, here $H_0=\hbar\omega/(g\mu_B)$ is a
paramagnetic resonance field for the given $g$-factor. At low
temperatures intensities of these components are different,
component corresponding to the transition from low-energy sublevel
dominates. At $\vect{H}||z$, left component ($H=H_0-|D|/(g\mu_B)$)
is more intense at $D<0$, while right component dominates at
$D>0$. Additionally, a characteristic "two-quantum" absorption
corresponding to $\left| S_z=+1\right \rangle \rightarrow \left|
S_z=-1\right \rangle$ transition appears in the field
$H_{2q}=H_0/2$. This transition is forbidden in the axial case but
it becomes allowed if magnetic field deviates from the anisotropy
axis or if $E\neq 0$. The "two-quantum" absorption is a
fingerprint of an $S=1$ magnetic center.

\section{Experimental results and discussion}
\begin{figure}
 \centering
 \epsfig{file=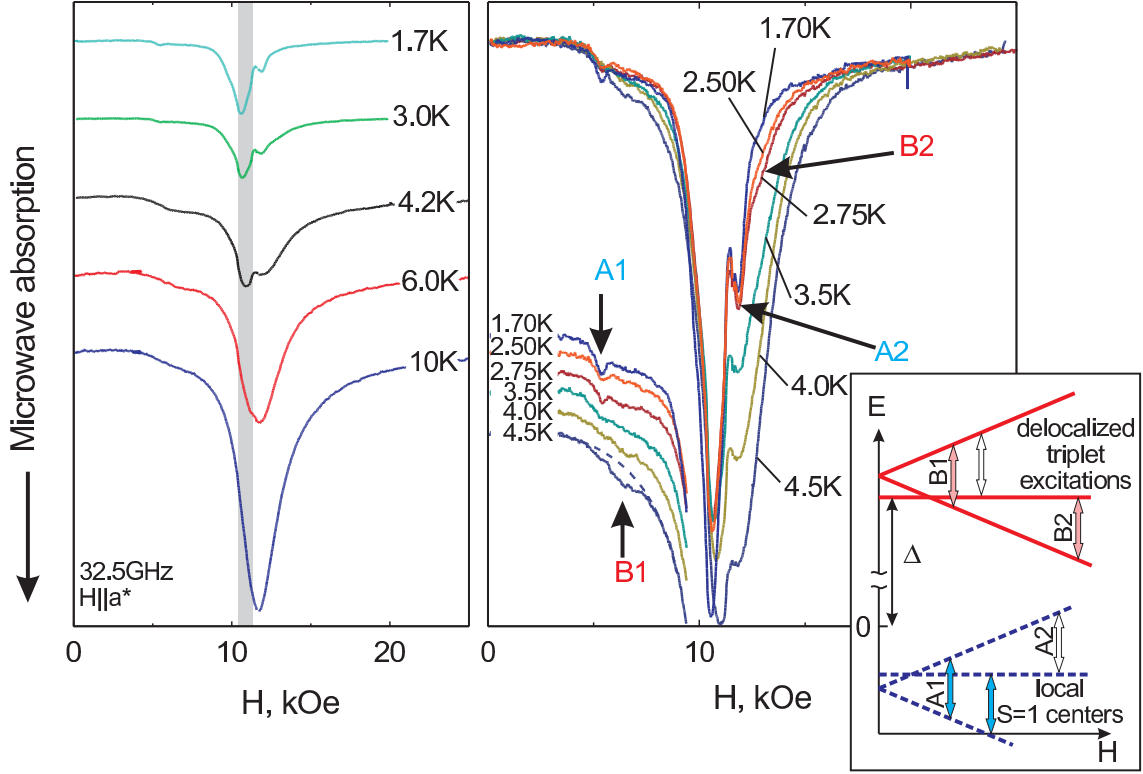, width=\figwidth, clip=}
 \caption{Temperature dependence of the ESR absorption in a
 Br-substituted PHCC with nominal Br concentration
 $x=7.5$\%. Left panel: representative data in the broad
 temperature range. Shaded area marks position of absorption due to uncontrolled defects as estimated from 3.0K data. Right panel: selection of low-temperature
 (below 4.5K) ESR absorption curves illustrating development of
 ESR response from local $S=1$ centers. Components of the ESR absorption
 related to local $S=1$ centers are denoted as "A1" and "A2", components of
 the ESR absorption corresponding to gapped triplet excitations
 are denoted as "B1" and "B2". Inset to the right panel: scheme of the
 zero-field splitting of triplet state for the triplet
 excitations (solid lines) and for the local $S=1$ centers (dashed),
 position of the singlet state is not shown, magnetic field is
 assumed parallel to the anisotropy axis for both sorts of
 triplets. Broad vertical arrows marks possible ESR transitions, filled arrows corresponds to the transitions
 from the lowest sublevel. ESR transitions are marked as the corresponding absorption components.}\label{fig:scansht}
\end{figure}

\begin{figure}
 \centering
 \epsfig{file=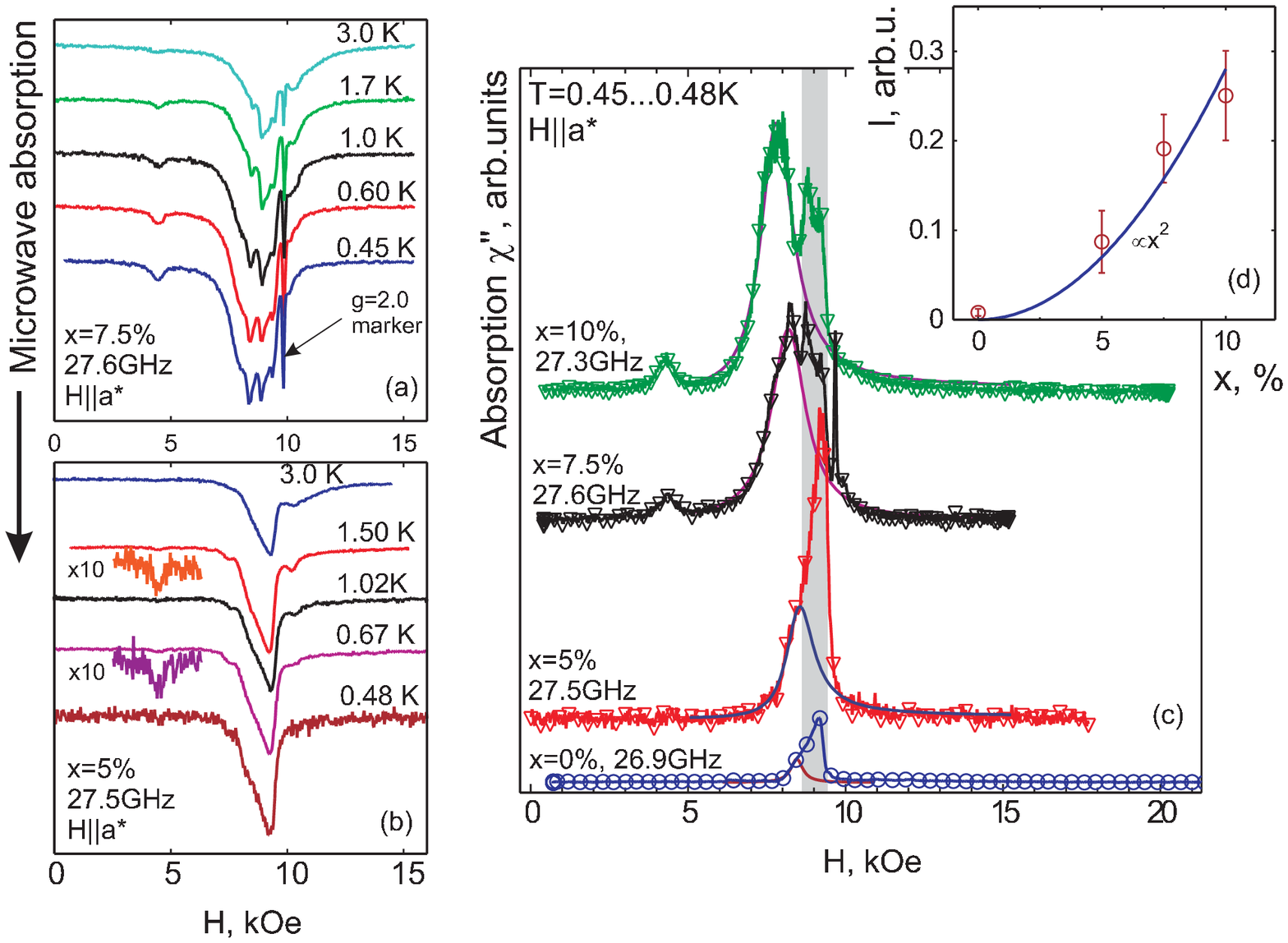, width=\figwidth, clip=}
 \caption{Low-temperature ESR absorption in Br-substituted PHCC.
 (a),(b) Representative ESR absorption spectra for the samples
 with nominal Br concentration $x=5,~7.5$\%. (c) Comparison of
 the scaled microwave absorption at the base temperature of the
 cryostat ($T=0.45\ldots0.48$K) for different bromine
 concentration. Lines with symbols are experimental curves, solid
 lines are Lorentzian fits. Scaling procedure is described in the
 text. Data are offset for better presentation. Shaded area marks
 ESR absorption by the uncontrolled defects which was excluded
 from the fit procedure. (d) Concentration dependence of the
 intensity of the impurity-induced absorption.}\label{fig:scanslt}
\end{figure}

\subsection{Summary of the earlier results.}

Energy spectrum of the doped PHCC was studied on the samples from
the same set as the samples earlier studied by means of inelastic neutron
scattering \cite{prb-dan-doped,prb-dan-doped-magnons}. INS study
demonstrates that excitation spectrum of the bromine substituted
PHCC remains gapped, energy gap increases with doping being equal
to 1 meV for the pure PHCC and 1.5 meV for
$x=7.5$\% \cite{prb-dan-doped}.

ESR absorption in pure and bond-diluted PHCC was reported in
Refs. \cite{phcc-pure,phcc-doped}. Magnetic resonance
response of a spin-gap magnet is a combination of a signal from
triplet excitations and paramagnetic defects.
These two contributions can be resolved due to their characteristic
temperature dependences.

At temperatures above 10K magnetic resonance absorption spectra in
pure and doped PHCC is dominated by a Lorentzian line with
anisotropic $g$-factor slightly above 2.0, as is typical for a
Cu$^{2+}$ ion. On cooling below 10K intensity of the ESR
absorption decreases as triplet excitations freeze out. At low
temperatures (below approximately 4K) gas of triplet excitations
is diluted, effects of quasiparticles interaction are switched off
and ESR absorption spectrum transforms into a multi-component
absorption typical for the $S=1$ object in the presence of
anisotropy. This spectrum consists of two intense components
shifted to the left and to the right from the higher temperature
resonance position and of one weaker component at approximately
half of the higher temperatures resonance field ("two-quantum"
absorption). For the pure PHCC these split components continue to
lose intensity on cooling and ESR absorption at lowest
temperatures consists of paramagnetic impurities (termed uncontrolled defects for the reminder of the paper) response
only \cite{phcc-pure}.

In the case of strongly doped PHCC with nominal bromine
concentration $x=10$\% low temperature evolution of ESR absorption
was quite distinct. ESR absorption typical for a $S=1$ object
was observed down to 500 mK.
Temperature dependence of the intensity of this low-temperature
absorption was described assuming presence of approximately 0.4\%
of gapless $S=1$ centers (gapless triplets) per molecule of PHCC
 \cite{phcc-doped}. These gapless triplets were interpreted as
localized triplet modes caught in the potential well created by
double substitution of a chlorine ions by bromine in a dimer.

While ESR absorption spectra of $x=5$\% Br-substituted sample
above 1.7K have demonstrated similar tendency \cite{phcc-doped},
there were no low-temperature (below 1 K) data to prove it
unambiguously. Also data for an intermediate bromine concentration
were missing. In the following sections we will close these
lacunas.

\subsection{Coexistence of ESR absorption from local $S=1$ centers and
delocalized triplets.}

Firstly, we will demonstrate here that ESR absorption spectra of the
intermediately bromine-substituted sample with $x=$7.5\% provides
evidences for coexistence of the triplet absorption from the
gapped delocalized triplets and local $S=1$ centers over a certain temperature range. Such coexistence indicates that gapped triplets and local $S=1$ centers are actually different objects decoupled at low temperatures.

Evolution of the ESR absorption in the 7.5\% nominally substituted
sample is presented in Figure \ref{fig:scansht}. At 10K main
absorption signal is almost a single-component paramagnetic resonance
with remnants of the "two-quantum" absorption visible. On cooling
ESR absorption shape became distorted. At 4.2K "two-quantum"
absorption becomes more pronounced, irregularly shaped paramagnetic
absorption appears and main absorption signal broadens and splits. Boundaries of paramagnetic resonance absorption due to uncontrolled defects can be estimated from low temperature data, this signal has quite distinguishable powder-like shape (see also data for nominally pure sample in Figure \ref{fig:scanslt} for comparison). These boundaries are shaded in Figure \ref{fig:scansht}.
On further cooling components of the main absorption signal
continue to loose intensity and around 2K they are replaced by
another more narrow signal. The same evolution can be traced for
the "two-quantum" transition.

At the intermediate temperatures coexistence of these two sorts of
signals can be followed (see right panel of Figure
\ref{fig:scansht}): with cooling narrow components (labeled as
"A1" and "A2") gain intensity and broad components (labeled as "B1" and "B2") lose
intensity. Both sorts of signals correspond to the $S=1$ objects,
"two-quantum" transition being the main fingerprint. Signals "B1" and "B2"
disappearing with cooling are naturally related to the gapped
delocalized triplet excitations. Signal that gains intensity with
cooling is due to the local $S=1$ centers.

At the lowest temperature of 1.7K spectral weight is shifted to the left from the
paramagnetic resonance position. This is in agreement with earlier
observations on pure and 10\% nominally doped PHCC: for the pure
compound (delocalized triplets only) right component of the split
signal is more intense, while low-temperature absorption in the
10\% doped sample (which was due to the local $S=1$ centers) was
dominated by its left component.

This difference in the position of more intense component correspond to different sign of effective anisotropy constant for the delocalized triplets and local $S=1$ centers. Note, that there is no single-ion anisotropy effects for the $S=1/2$ copper ions of PHCC --- thus, these effective anisotropy constants both for the delocalized excitations and for the local $S=1$ centers are due to some anisotropic spin-spin interactions.

\subsection{Concentration dependence of the low-temperature
absorption}

Here we will compare low-temperature absorption from different
samples below 1K, when all absorption is due to the
uncontrolled defects and local $S=1$ centers.

Low-temperature ESR absorption spectra are shown in Figure
\ref{fig:scanslt}. We observe several absorption components: characteristic powder-like absorption due to uncontrolled defects, split absorption due to the $S=1$ centers and weaker "two-quantum" absorption signal at approximately half of the paramagnetic resonance field. Uncontrolled paramagnetic impurities yield irregularly shaped
absorption signal at almost the same fields, which complicates
extraction of the localized triplets response. However, boundaries of this absorption are approximately the same (at the same microwave frequency) and can be reliably established from the analysis of all data (see shaded area in Figure \ref{fig:scanslt}-c).

To compare absorption curves measured on different samples we have extracted
imaginary part of the susceptibility and scaled its integral at temperature of 3-5K to
the known static susceptibility data. This procedure allowed to recalculate absorption for different samples to the same scale. Scaled curves demonstrate
regular increase of absorption to the left of the paramagnetic
resonance field with increase of bromine concentration. This absorption corresponds to one of the split components of local $S=1$ centers.

To analyze intensity of the split components we have to separate contribution from uncontrolled defects, which is clearly complicated at low bromine content. Weaker
"two-quantum" signal is well separated from the paramagnetic
impurities signal, however it is impossible to compare its intensity in
different samples as intensity of this "almost forbidden"
transition strongly depends on excitations conditions which were
not reproducible. Absorption in the vicinity of paramagnetic resonance on the
contrary is excited by usual transverse polarization which allows
to scale it with static magnetization data. To exclude
uncontrolled impurities signal we simply ignored data in its
vicinity (shaded area in Figures \ref{fig:scansht} and \ref{fig:scanslt}) during the fit.

As a result of the fit intensities of the ESR absorption due to
the localized triplets were found. Accuracy of intensity
determination is about 20\% being hindered by uncertainties of the
scaling procedure and some arbitrariness in the paramagnetic
signal exclusion. To check the reliability of this procedure we made similar fit for the nominally pure sample, which yielded possible contribution tenfold smaller than for the $x=5$\% sample.

Intensity of the split component is proportional to the
concentration of $S=1$ objects. It approximately follows $x^2$ dependence
thus supporting the model of double substitution in a dimer
suggested in Ref. \cite{phcc-doped}.

Position of the split component to the left from paramagnetic resonance position again corresponds to the inversion of effective anisotropy constant sign for the local $S=1$ centers as compared to triplet excitations (see energy levels scheme at Figure \ref{fig:scansht}). Magnitude of the shift from the paramagnetic resonance position is slightly concentration dependent: splitting increases with increasing bromine content. Variation of this splitting with bromine content is not unexpected: doping slightly modifies lattice parameters \cite{dan} which could lead to modification of anisotropic spin-spin couplings.

\section{Conclusions}

We have demonstrated that $S=1$ centers are formed in a spin-gap magnet PHCC with non-magnetic bond-doping. These centers are presumably formed from a couple of $S=1/2$ Cu$^{2+}$ ions on double substitution of intra-dimer chlorine by bromine. We have shown that at low temperatures these centers are decoupled from the triplet excitations and that the effective anisotropy constant for local $S=1$ centers differs from that for the triplet excitations.

The work was supported by Russian Foundation for Basic Research
Grant No.15-02-05918. This work was partially supported by
the Swiss National Science Foundation, Division 2 and by the Estonian Ministry of Education and Research under grant No. IUT23-03 and the Estonian Research Council grant No. PUT451.

\end{document}